\documentclass[sigconf]{acmart}

\usepackage{stackengine}
\usepackage{booktabs} 
\usepackage{url}
\usepackage{color}
\usepackage{soul}
\usepackage{booktabs}
\usepackage{graphicx}
\usepackage{multirow}
\usepackage{mathrsfs}
\usepackage{amssymb}
\usepackage{amsbsy}
\usepackage{amsmath}
\usepackage{tcolorbox}
\usepackage{balance}
\usepackage[flushleft]{threeparttable}
\usepackage{flushend}
\usepackage{listings}
\usepackage{algorithm}
\usepackage{algpseudocode}
\usepackage{esvect}

\algrenewcommand{\algorithmiccomment}[1]{/* #1 */}

\newlength\llength
\definecolor{mygreen}{rgb}{0,0.6,0}
\definecolor{mygray}{rgb}{0.5,0.5,0.5}
\definecolor{mymauve}{rgb}{0.58,0,0.82}

\global\long\def\hb{\boldsymbol{h}}

\global\long\def\ib{\boldsymbol{i}}
\global\long\def\fb{\boldsymbol{f}}

\global\long\def\ob{\boldsymbol{o}}

\global\long\def\cb{\boldsymbol{c}}
\global\long\def\bb{\boldsymbol{b}}

\global\long\def\wb{\boldsymbol{w}}

\global\long\def\hb{\boldsymbol{h}}

\global\long\def\bb{\boldsymbol{b}}
\global\long\def\cb{\boldsymbol{c}}

\global\long\def\fb{\boldsymbol{f}}
\global\long\def\ib{\boldsymbol{i}}

\setcopyright{rightsretained}

\acmDOI{XXX}

\acmISBN{XXX}

\acmConference[XXX]{XXX}{XXX}{XXX}
\acmYear{2018}
\copyrightyear{2018}

\acmPrice{15.00}

\begin{document}
	\title{A deep tree-based model for software defect prediction}
	

\author{Hoa Khanh Dam}
\affiliation{%
	\institution{University of Wollongong, Australia}
}
\email{hoa@uow.edu.au}

\author{Trang Pham}
\affiliation{%
	\institution{Deakin University, Australia}
}
\email{phtra@deakin.edu.au}

\author{Shien Wee Ng}
\affiliation{%
	\institution{University of Wollongong, Australia}
}
\email{swn881@uowmail.edu.au}

\author{Truyen Tran}
\affiliation{%
	\institution{Deakin University, Australia}
}
\email{truyen.tran@deakin.edu.au}

\author{John Grundy}
\affiliation{%
	\institution{Deakin University, Australia}
}
\email{j.grundy@deakin.edu.au}

\author{Aditya Ghose}
\affiliation{%
	\institution{University of Wollongong, Australia}
}
\email{aditya@uow.edu.au}

\author{Taeksu Kim}
\affiliation{%
	\institution{Samsung Electronics, Republic of Korea}
}
\email{taeksu.kim@samsung.com}

\author{Chul-Joo Kim}
\affiliation{%
	\institution{Samsung Electronics, Republic of Korea}
}
\email{chuljoo1.kim@samsung.com}

	
	\renewcommand{\shortauthors}{Author et al.}

	\begin{abstract}
    Defects are common in software systems and can potentially cause various problems to software users. Different methods have been developed to quickly predict the most likely locations of defects in large code bases. Most of them focus on designing features (e.g. complexity metrics) that correlate with potentially defective code. Those approaches however do not sufficiently capture the syntax and different levels of semantics of source code, an important capability for building accurate prediction models. In this paper, we develop a novel prediction model which is capable of automatically learning features for representing source code and using them for defect prediction. Our prediction system is built upon the powerful deep learning, tree-structured Long Short Term Memory network which directly matches with the Abstract Syntax Tree representation of source code. An evaluation on two datasets, one from open source projects contributed by Samsung and the other from the public PROMISE repository, demonstrates the effectiveness of our approach for both within-project and cross-project predictions.	
	\end{abstract}
	
	%
	%



	\maketitle
	
    \section{Introduction}

As software systems continue playing a critical role in all areas of our society, defects arisen from those software have significant impact onto businesses and people's lives. Identifying defects in software code however becomes increasingly difficult due to the significant grow of software codebase in both size and complexity. The importance and challenges of defect prediction have made it an active research area in software engineering. Substantial research have gone into developing predictive models and tools which help software engineers and testers to quickly narrow down the most likely defective parts of a software codebase \cite{KameiS16,D'Ambros:2012:EDP,Catal:2009:SRS}. Early defect prediction helps prioritize and optimize effort and costs for inspection and testing, especially when facing with cost and deadline pressures.

Machine learning techniques have been widely used to build defect prediction models. Those techniques derive a number of features (i.e. predictors) from software code and feed them to common classifiers such as Naive Bayes, Support Vector Machine and Random Forests. Substantial research (e.g. \cite{Hall:2012,Moser:2008,Nagappan:2005:URC,Pinzger:2008:DNP,Nagappan:2008:IOS,Hassan:2009:PFU}) have gone into carefully designing features which are able discriminate defective code from non-defective code such as code size, code complexity (e.g. Halstead features, McAbe, CK features, MOOD features), code churn metrics (e.g. the number of code lines changed), process metrics. However, those features do not truly reflect the syntax and semantics of code. In addition, software metric features normally do not generalize well:  features that work well in a  certain  software  project  may  not  perform  well  in  other projects \cite{Zimmermann:2009:CDP}.

Natural Language Processing techniques have also been leveraged to extract defect predictors from code tokens in source files. A common technique is using Bag-of-Words (BoW) which treats code tokens as terms and represents a source file as term-frequencies. The BoW approach is however unable to detect differences in the semantics of source code due to differences in code order or syntactic structure (e.g. $x \ge y$ vs. $y \ge x$).  Hence, recent trends started to focus on persevering code structure information in representing source code. However, recent work such as \cite{Wang:2016:ALS} does not fully encode the syntactic structure of code nor the semantics of code tokens, e.g. fails to recognize the semantic relations between ``for'' and ``while''.

This paper presents a novel deep tree-based model for defect prediction. We leverage Long Short-Term Memory (LSTM) \cite{hochreiter1997long}, a powerful deep learning architecture to capture the long context relationships in source code where dependent code elements are scattered far apart. The syntax and different levels of semantics in source code are usually represented by tree-based structures such as Abstract Syntax Trees (ASTs). Hence, we adapted a tree-structured LSTM network \cite{Tai-TreeLSTM2015} in which the LSTM tree in our prediction system matches exactly with the AST of an input source file, i.e. each AST node corresponds to an LSTM unit in the tree-based network. The contributions of our paper are as below.

\begin{enumerate}

  \item A deep tree-based LSTM model for source code which effectively preserve both syntactic and structural information of the programs (in terms of ASTs). Through an AST node embedding mechanism, our representation of code tokens also preserve their semantic relations. 
 
 \item A prediction system which takes as input a ``raw'' Abstract Syntax Tree representing a source file and predict if the file is defective or clean. The features are automatically learned through the LSTM model, thus eliminating the need for manual feature engineering which occupies most of the effort in traditional approaches.



  \item  An extensive evaluation using real open source projects provided by Samsung and the PROMISE repository\footnote{http://openscience.us/repo/} demonstrates the empirical strengths of our model for defect prediction. 
\end{enumerate}

The outline of this paper is as follows. In the next section, we provide a motivation example, followed by an overview of our approach in Section \ref{sect:approach}. Section \ref{sect:model-building} describes how our prediction model is built. We then describe how the model is trained and implemented in Section \ref{sect:model-training}. We report a number of experiments to evaluate our approach in Section \ref{sect:eval}. In Section \ref{sect:related-work}, we discuss related work before summarizing the contributions of the paper and outlines future work in Section \ref{sect:conclusions}.

    \section{Motivating example}

We start with an example which illustrates the challenges when using existing approaches for software prediction. Figure \ref{fig:example} shows two simple code listings written in Java. Both contains a \emph{while} loop in which the integer at the top of a given \emph{stack} is repeatedly removed through the \emph{pop} operation. Listing 1 has a defect: if the given stack's size is smaller than 10, underflow exception can occur when the stack is empty and the \emph{pop} operation is executed. Listing 2 rectifies this issue by checking if the stack is not empty just before invoking the \emph{pop} operation.

\setbox0=\hbox{%
\begin{minipage}{1.5in}
\begin{lstlisting}[caption=A.java]
int x = 0;
if (!stack.empty()) {
  while (x < 10) {
    int y;
    y = stack.pop();
    x++;
  }
}
\end{lstlisting}
\end{minipage}
}
\savestack{\listingA}{\box0}

\setbox0=\hbox{%
\begin{minipage}{1.5in}
\begin{lstlisting}[caption=B.java]
int x = 0;
while (x < 10) {
  int y;
  if (!stack.empty()) {
    y = stack.pop();
  }
  x++;
}

\end{lstlisting}
\end{minipage}
}
\savestack{\listingB}{\box0}

\begin{figure}[ht]
    \begin{tabular}{|c|c}
    \listingA &
    \listingB \\
    \end{tabular}
    \caption{A motivating example}
	\label{fig:example}
\end{figure}

In the above example, existing techniques for defect prediction would face the following challenges:

\begin{enumerate}
  \item \textbf{Similar software metrics}: The two code listings are identical with respect to the number of code lines, conditions, variables, loops, and branches. Thus, they would be indistinguishable if software metrics (as widely used in existing approaches \cite{Hall:2012}) are used as features. In may other cases,  two pieces of code may have the same metrics but they behave differently and thus have different likelihood of defectiveness.

  \item \textbf{Similar code tokens and frequencies}: Recent approaches looked into the actual code content and represent a source code file as a collection of code tokens (e.g. \emph{int}, \emph{x}, \emph{if}, etc.) associated with frequencies (e.g. 2 for \emph{int} in Listing 1). The Term-frequencies are then used as the predictors for defect prediction. However, this is not necessarily the best presentation for code. In fact,  the code tokens and their frequencies are also identical in both code listings. Hence, relying only on the term-frequency features would fail to recognize that Listing 1 has a defect while the Listing 2 does not.

  \item \textbf{Syntactic and semantic structure}: The two code listings are different in their structure and thus would behave differently. The location of the \emph{if statement} makes a significant difference in causing or removing a defect. Syntactic structure also requires pairs of code element appear together (e.g., \emph{try} and \emph{catch} in Java, or file open and close). $n$-\emph{grams} models are commonly used to capture those repetitive sequential patterns in code. However, $n$-\emph{grams} models are usually restricted to a few code elements, thus are insufficient for cases where dependent code elements scatter far apart. In addition, code elements are not always required to follow a specific order, e.g. in code listing 1, lines 5 and 6 can be swapped without changing the code's behaviour.

  \item \textbf{Semantic code tokens}: Code elements has their own semantics. For example, in Java ``for'' and ``while'' are semantically similar, e.g. the while loop in the above code listings can be replaced with a for loop without changing the code behaviour. Existing approaches (e.g. \cite{Yang:2015:DLJ,Wang:2016:ALS}) often overlook those semantics of code tokens.

\end{enumerate}

The syntax and different levels of semantics in source code are usually represented by tree-based structures such as Abstract Syntax Trees (ASTs). Hence, to address the above challenges, we develop a deep tree-based LSTM neural network to model the Abstract Syntax Trees of source code. This representation effectively preserves both syntactic and structural information of the code, and thus is used for defect prediction.

\begin{figure*}[ht]
\centering \includegraphics[width=0.9\linewidth]{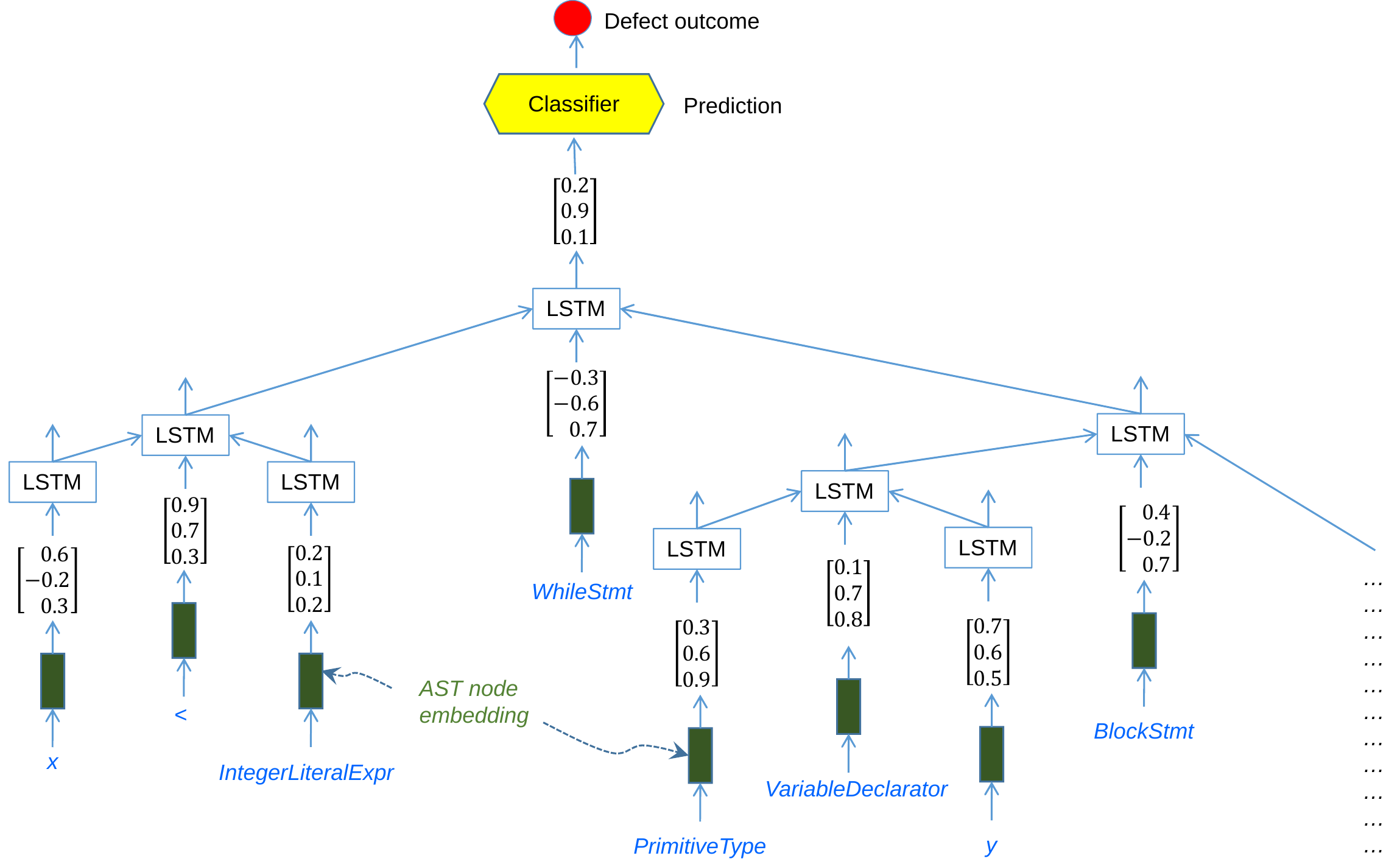}
\caption{An example of how a vector representation is obtained for a code sequence}
\label{fig:tree-LSTM-example}
\end{figure*}

\section{Approach}\label{sect:approach}

Most of existing work in defect prediction focus on determining whether a source file is likely to be defective or not. This level of granularity has become a standard in the literature of software defect prediction. Determining if a source file is defective can be considered as a function $predict(f)$ which takes as input a file $f$ and returns either 1 for defective and 0 for clean. We approximate this classification function $predict(x)$ ((or also referred to as the model) by learning from a number of examples (i.e. files known to be defective or clean) provided in a \emph{training set}.

Our prediction model is built upon the Long-Short Term Memory, a powerful deep learning architecture. Unlike existing work, our model is constructed as a tree-structured network of LSTM units to better reflect the syntactic and many levels of semantics in source code. After training, the learned function  is used to automatically determine the defectiveness of new files in the same project (within-project prediction) or in a different project (cross-project prediction). Through employing a novel attention mechanism into our tree-based LSTM network, our model is also able to locate the parts (e.g. code lines) in a source file that are likely the cause of a defect. This helps understand and diagnose exactly what the model is considering and to what degree for specific defects. The key steps of our approach (see Figure \ref{fig:tree-LSTM-example}) is as below.

\begin{enumerate}
  \item Parse a source code file into an Abstract Syntax Tree (see Section \ref{subsec:Parsing-source-code} for details).
  \item Map AST nodes to continuous-valued vectors called embeddings (Section \ref{subsec:embedding}).
  \item Input the AST embeddings to a tree-based network of LSTMs to obtain a vector representation of the whole AST. Input this vector to a traditional classifier (e.g. Logistic Regression or Random Forests) to predict defect outcomes (Section \ref{subsec:main-model}).
\end{enumerate}

In the next section, we will describe each of these steps in details.

\section{Model building}\label{sect:model-building}

\subsection{Parsing source code \label{subsec:Parsing-source-code}}


We parse each source code file into an Abstract Syntax Tree (AST). This process ignored comments, blank lines, punctuation and delimiters (e.g. braces, semicolons, and parentheses). Each node of the AST represents a construct occurring in the source code. For example, the root of the AST represents a whole source file, and its children are all the top element of the file such as import and class declarations. Each class declaration node (i.e. \texttt{ClassOrInterfaceDeclaration}) has multiple children nodes which represent the fields (\texttt{FieldDeclaration}) or the methods (\texttt{MethodDeclaration}) of the class. A method declaration node also has multiple children nodes which represent its name, argument parameters, return type, and body.

\begin{figure}[ht]
\centering \includegraphics[width=0.8\linewidth]{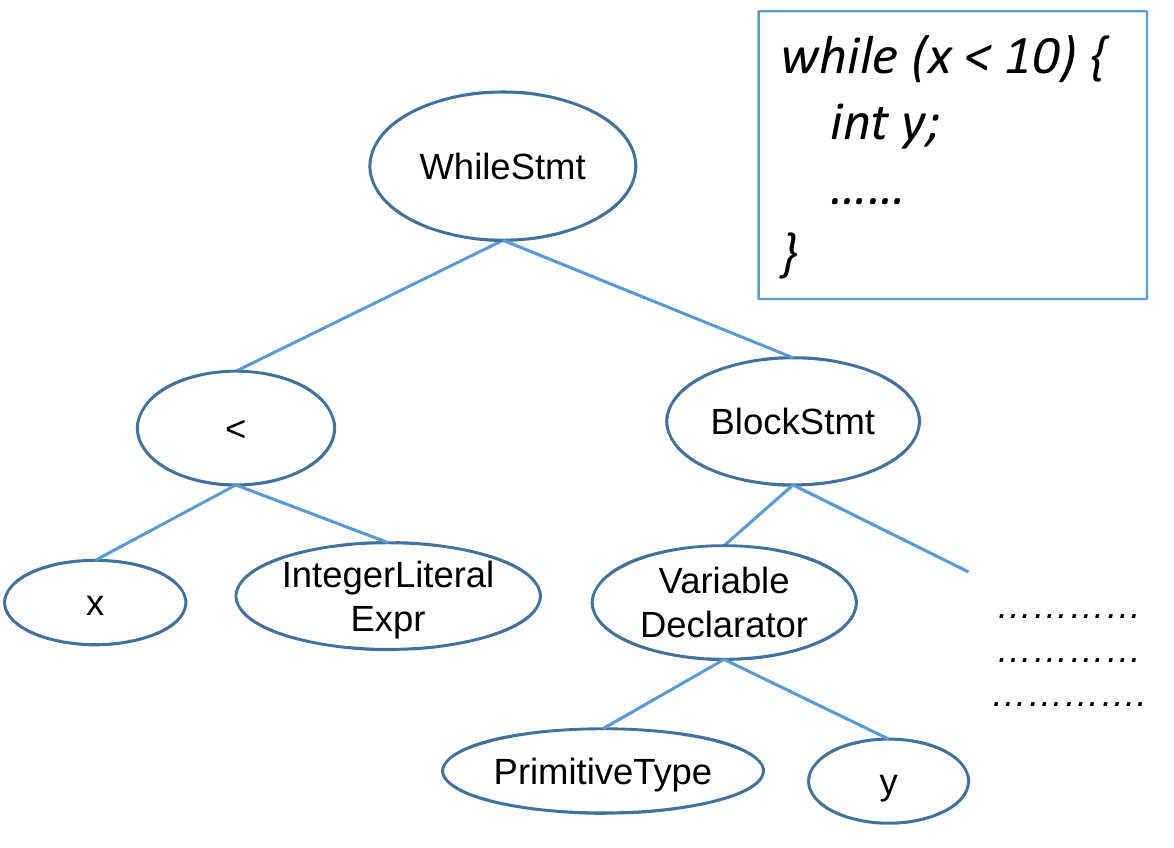}
\caption{An example of an Abstract Syntax Tree (AST) for a Java program}
\label{fig:AST-example}
\end{figure}

We label each tree node with its AST type (e.g. \texttt{FieldDeclaration}, \texttt{MethodDeclaration}, \texttt{BlockStmt}, and \texttt{WhileStmt}) or its AST name (e.g. variable name, class name, and method name) in the case of SimpleName nodes  (see Figure \ref{fig:AST-example}). Constant integers, real numbers, exponential notation, hexadecimal numbers and strings are represented as AST nodes of their type (rather than the actual number or string) since they are specific to a method or class. For example, the integer number 10  is represented as a \texttt{IntegerLiteralExpr} node (see Figure \ref{fig:AST-example}), while a string ``Hello World'' is represented as a \texttt{StringLiteralExpr}.

The unique label names collected from all AST tree nodes in the entire corpus are used to form a vocabulary. Following standard practice (e.g. as done in \cite{White:2015:TDL}), we also replace less popular tokens (e.g. occurring only once in the corpus) and tokens which exist in test sets but do not exist in the training set with a special token $\langle unk\rangle$. A fixed-size vocabulary $\mathscr{V}$ is constructed based on top $N$ popular tokens, and rare tokens are assigned to $\langle unk\rangle$. Doing this makes our corpus compact but still provides partial semantic information.

\subsection{Embedding AST nodes}\label{subsec:embedding}

Each AST node is input to an LSTM unit. Since the LSTM unit only takes input in the form of vectors, we need to map the label name of each AST node into a fixed-length continuous-valued vector. We refer to this AST node embedding process as \emph{ast2vec}.

This process makes use of an embedding matrix $\mathcal{M}\in\mathbb{R}^{d\times|\mathscr{V}|}$ where $d$ is the size of a AST node embedding vector and $|\mathscr{V}|$ is the size of vocabulary $\mathscr{V}$. Each AST node label has an index in the vocabulary (i.e. encoded as one-hot vector).  The embedding matrix acts as a look-up table: an AST node label $i^{th}$ is mapped to column vector $i^{th}$ in  matrix $\mathcal{M}$. For example in Figure \ref{fig:tree-LSTM-example}, a \texttt{WhileStmt}  node is embedded in vector ${[}-0.3, -0.6, 0.7{]}$, while \texttt{IntegerLiteralExpr} is mapped to vector ${[}0.2, 0.1, 0.2{]}$. The embedding process offers two benefits. First, an embedding vector has lower dimensions than a one-hot vector (i.e. $d < |\mathscr{V}|)$). Second,  in the embedding space, AST nodes that frequently appear in similar context are close to each other. This often leads to code elements with similar semantic being neighbours. For example, the embeddings of \texttt{WhileStmt} and \texttt{ForStmt} would be close to each other in the embedding space.

\begin{figure}[ht]
\centering \includegraphics[width=0.8\linewidth]{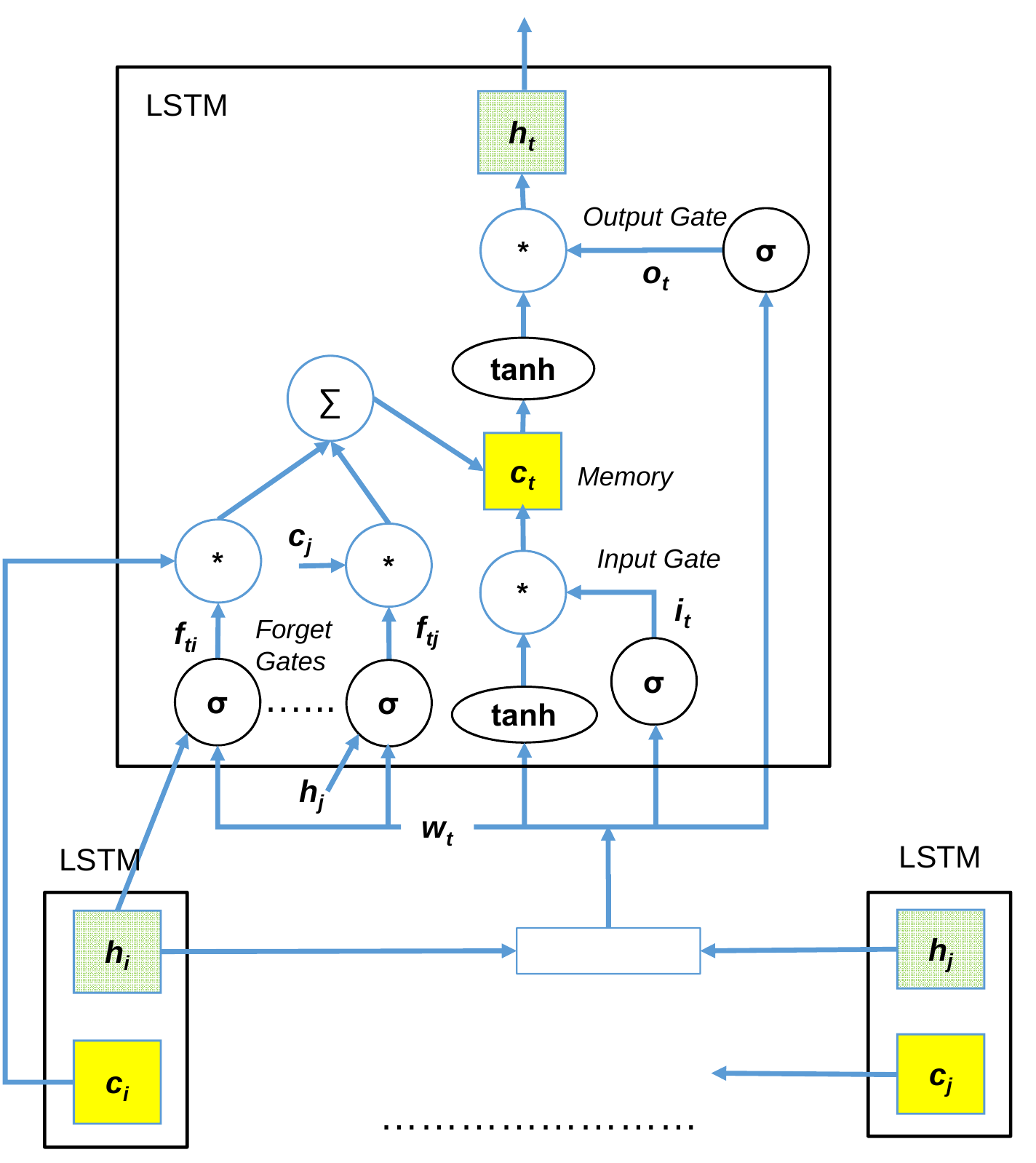}
\caption{The internal structure of an Tree-LSTM unit}
\label{fig:tree-LSTM-compose}
\end{figure}

The embedding matrix is randomly initialized, and then is adjusted as part of the training process, which we will discussed in Section \ref{sect:model-training}.


\subsection{Defect prediction model}\label{subsec:main-model}

Our prediction model is represented as function $predict()$ which takes as input a source file and returns 1 if the file is defective and 0 otherwise (see Algorithm \ref{algorithm:tree-lstm}). It first parses the source file into an Abstract Syntax Tree (line 2 in Algorithm \ref{algorithm:tree-lstm}). The root of the AST is fed into a Tree-LSTM unit to obtain a vector representation $\hb_{root}$ (line 3). This vector is fed into to a traditional classifier to compute the probability of the file being defective. If this probability is not smaller than 0.5, the function returns 1. Otherwise, it returns 0 (lines 4--6).


\begin{algorithm}
	\caption{Tree-based defect prediction. Model parameters include ($W_{for}, U_{for}, b_{for}$), ($W_{in}, U_{in}, b_{in}$), ($W_{ce}, U_{ce}, b_{ce}$), and ($W_{out}, U_{out}, \bb_{out}$) shared by all Tree-LSTM units.}
	\label{algorithm:tree-lstm}
	\begin{algorithmic}[1]

        \Function{predict}{File f}
            \State $root \gets parseFile2AST(f)$
    		\State $(\hb_{root}, \cb_{root}) \gets  t{\text -}lstm\left(root\right)$
            \State $\hat{p} \gets  classifier\left(\hb_{root}\right)$
            \If{$\hat{p} \ge 0.5$}
                \State \textbf{return} $1$  
             \Else
                \State \textbf{return} $0$  
             \EndIf
        \EndFunction

        \Statex


        \Function{t-lstm}{ASTnode t}
            \State $\wb_{t} \gets ast2vec(getNodeName(t))$

            \State $C(t) \gets getChildrenNodes(t)$

            \State $(\hb_{k}, \cb_{k}) \gets (\vv{0}, \vv{0})$ 		

            \ForAll {$ASTNode$ $k \in C(t)$}
    		      \State $(\hb_{k}, \cb_{k}) \gets t{\text -}lstm(k)$
                  \State $\fb_{tk}=sigmoid\left(W_{for}\wb_{t} + U_{for}{\hb_{k}} + b_{for}\right)$
            \EndFor

            \State $\tilde{\hb} \gets \sum\limits _{k \in C(t)}\hb_{k}$

            \State $\ib_{t} \gets sigmoid\left(W_{in}\wb_{t} + U_{in}\tilde{\hb} + b_{in}\right)$

    		\State $\tilde{\cb_{t}} \gets tanh\left(W_{ce}\wb_{t} + U_{ce}\tilde{\hb} + b_{ce}\right)$

             \State $\cb_{t}=\ib_{k} * \tilde{\cb_{t}} + \sum\limits _{k \in C(t)} \fb_{tk} * \cb_{k}$

             \State $\ob_{t}=sigmoid\left(W_{out}\wb_{t} + U_{out}\tilde{\hb} + b_{out}\right)$

             \State $\hb_{t}=\ob_{t} * tanh\left(\cb_{t}\right)$

             \State \textbf{return} $(\hb_{t}, \cb_{t})$
        \EndFunction
	\end{algorithmic}
\end{algorithm}

An Tree-LSTM unit (see Figure \ref{fig:tree-LSTM-compose}) is modeled as function $t{\text -}lstm()$, which takes as input an AST node $t$ and outputs two vectors: $\hb$ (representing the hidden output state) and $\cb$ (representing the context it remembers so far in the AST). This is done by aggregating those outputs from the descendants, i.e. calling $t{\text -}lstm()$ recursively on the children nodes (lines 11--26).  This function first obtains the embedding $\wb_{t}$ of the input AST node $t$ (using $ast2vec$ as discussed in Section \ref{subsec:embedding}). It then obtains all the children node $C(t)$ of node $t$, and each child node $k \in C(t)$ is fed into an LSTM unit to obtain the pair of hidden output state and context vectors $(\hb_{k}, \cb_{k})$ for each child node. These are then used to compute the pair of hidden output state and context vectors $(\hb_{t}, \cb_{t})$ for the parent node as follows.

How information embedded in $\wb_{t}$ and $(\hb_{k}, \cb_{k})$ (for all $k \in C(t)$) flow through an Tree-LSTM unit is controlled by three important components: an input gate (represented as $\ib_{t}$), an output gates ($\ob_{t}$) and a number of forget gates (one $\fb_{tk}$ for each child node $k$). These components depend on the input $\wb_{t}$ and the output state $\hb_{k}$ of the children. These correlations are encoded in groups of parameter matrices: ($W_{for}, U_{for}, b_{for}$) for the forget gates, ($W_{in}, U_{in}, b_{in}$) for the input gate, and ($W_{out}, U_{out}, \bb_{out}$) for the output gates.

A Tree-LSTM unit has a number of forget gates $\fb_{tk}$, one for each child node $k$ and is computed as a sigmoid function over $\wb_{t}$ and $\hb_{k}$ (line 17). A forget gate $\fb_{tk}$ has a value between 0 and 1, which enables the Tree-LSTM unit to selectively include information from each child. The output from children nodes  are combined to serve as an input the the parent LSTM unit (line 19). How much of these new information is stored in the memory cell is controlled by two mechanisms (lines 20--22). First, the input gate $\ib_{k}$, represented as a sigmoid function, controls decides which values will be updated. Second,  a vector of new candidate values $\tilde{\cb_{t}}$, which will be added to the memory cell, is created using a $tanh$ function.

The new memory is updated by multiplying  the old memory of each child by $\fb_{tk}$, leaving out the things we decided to forget earlier. We sum it over all the child node and then add this with $ \tilde{\cb_{t}}$. Finally, the output is a filtered version of the memory, which is controlled by the output gate $\ob_{t}$ (line 23).  We apply $tanh$ function to the memory (to scale the values to be between -1 and 1) and multiply it by the output of the sigmoid gate so that only some selected parts are output (line 24).

%
%
%
%
%
%

\section{Model training} \label{sect:model-training}

\subsection{Training Tree-LSTM  \label{subsec:Pre-training}}

\begin{figure}[ht]
\centering \includegraphics[width=\linewidth]{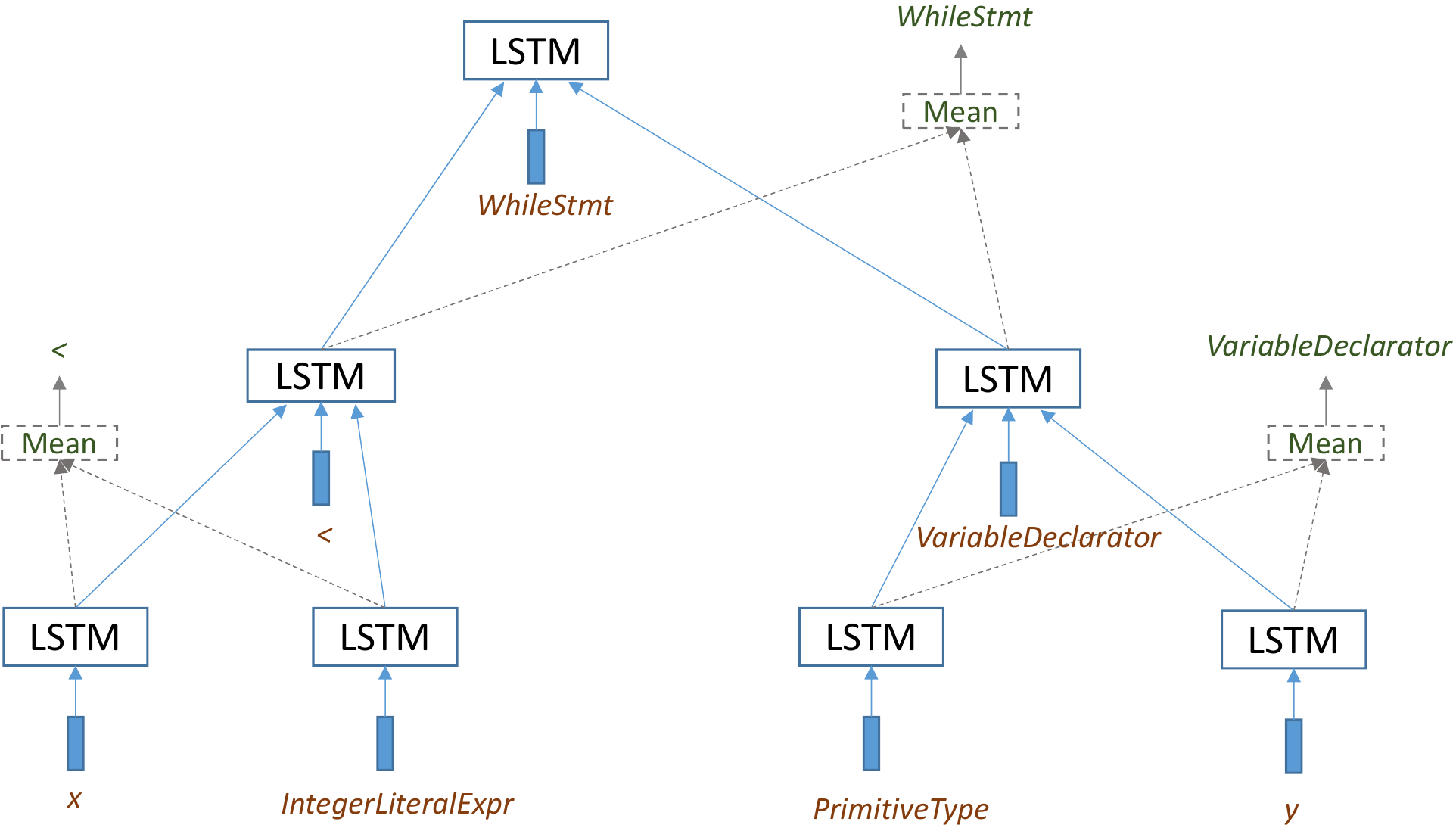}
\caption{Training Tree-LSTM by predicting the label name of a parent node from its children nodes}
\label{fig:training-tree-LSTM}
\end{figure}

We train the Tree-LSTM unit in a unsupervised manner, i.e. \emph{not} using the ground-truth defect labels. We leverage the strong predictiveness of AST, i.e. if we know the label name of all the children, we can predict the label name of its parent. Using a large number of AST branches, we train the Tree-LSTM unit through making such a prediction.
For example, the parent of ``<'' and ''VariableDeclarator'' is ``WhileStmt'', while the parent of ``x'' and ``IntegerLiteralExpr'' is ``<'' (see Figure \ref{fig:training-tree-LSTM}).

Specifically, each AST node $w_{t}$ has a set of children $C(t)$, and each $c_k \in C(t)$ has an output state $h_k$. We can predict the label name of the parent node using all its children hidden states through the softmax function:

\begin{equation}
P\left(w_{t}=w\mid w_{c1..ck}\right)=\frac{\text{exp}\left(U_{t}\tilde{h}_{t}\right)}{\sum_{w'}\text{exp}\left(U_{w'}\tilde{h}_{t}\right)}
\end{equation}

where $U_k$ is a free parameter and $\tilde{h}_{t}=\frac{1}{|C(t)|}\sum_{k=1}^{|C(t)|}h_{k}$

Let $\theta$ be the set of all parameters in the LSTM unit, which includes the embedding matrix $\mathcal{M}$ and weight matrices ($W_{for}, U_{for}, b_{for}$), ($W_{in}, U_{in}, b_{in}$), ($W_{ce}, U_{ce}, b_{ce}$), and ($W_{out}, U_{out}, \bb_{out}$). These parameters are initialized randomly and then learned through a training process. Training is involves three main steps: (i) input a AST branch in the training data to the LSTM units to obtain a prediction for the label name of the parent node in that branch; (ii) compare the difference $\delta$ between the predicted outcome and the actual outcome; (iii) adjusting the values of the model parameters such that the difference $\delta$ is minimized. This process is done iteratively for all files in the training data.

To measure the quality of a specific set of values for the model parameters, we define a loss function $L(\theta)$ which is based on the difference $\delta$ between the predicted outcome and the actual outcome. A setting of the model parameters $\theta$ that produces a correct prediction (e.g. the label name of a parent node is correctly predicted) would have a very low loss $L$. Hence, learning is achieved through the optimization process of finding the set of parameters $\theta$ that minimizes the loss function.

Since every component in the model is differentiable, we employ the widely-used stochastic gradient descent to perform optimization.  The optimization process is done through backpropagation: the model parameters $\theta$ are updated in the opposite direction of the gradient of the loss function $L(\theta)$. A learning rate $\eta$ is used to control how fast or slow we will move towards the optimal parameters. A large learning rate may miss the optimal solution, while a small learning rate will take too many iterations to converge to an optimal solution. We use RMSprop, an adaptive stochastic gradient method (unpublished note by Geoffrey Hinton), which is known to work best for recurrent models. We tuned RMSprop by partitioning the data into mutually exclusive training, validation, and test sets and running multiple training epoches. Specifically, the training set is used to learn a useful model. After each training epoch, the learned model was evaluated on the validation set and its performance was used to assess against hyperparameters (e.g. learning rate in gradient searches). Note that the validation set was \emph{not} used to learn any of the model's parameters. The best performing model in the validation set was chosen to be evaluated on the test set. We also employed the early stopping strategy, i.e. monitoring the model's performance during the validation phase and stopping when the performance got worse.

We have also implemented \emph{dropout} into our model \cite{srivastava2014dropout}, an effective mechanism to prevent overfitting in neural networks. Here, the elements of input and output states are randomly set to zeros during training. During testing, parameter averaging is used. In effect, dropout implicitly trains many models in parallel, and all of them share the same parameter set. The final model parameters represent the average of the parameters across these models. Typically, the dropout rate is set at $0.5$. We implemented the model in Theano \cite{Theano} and Keras\cite{Keras} frameworks, running in Python.  Theano supports automatic differentiation of the loss function and a host of powerful adaptive gradient descent methods. Keras is a wrapper making model building much easier. We employed Noise-Contrastive Estimation \cite{gutmann2012noise} to compute the softmax function . We also run multiple epoches against a validation set to choose the best model. We use \emph{perplexity}, a common intrinsic evaluation metric based on the log-loss, as a criterion for choosing the best model and early stopping.

\subsection{Training defect prediction model}

The above process enables us to automatically generate features for all the source files in the training set. These files with their features and labels (i.e. defective or clean) are then used to train machine learning classifiers by learning from a number of examples (i.e. files known to be defective or clean) provided in a \emph{training set}. We tried two alterative classifiers: Logistic Regression and Random Forests. Logistic Regression uses the logistic function (also called the sigmoid function) to approximate the probability of a source file being defective given its AST feature vector representation. Random Forests (RFs) \cite{breiman2001random} is a randomized ensemble method which combines the estimates from many decision trees to make a prediction.

    \section{Evaluation}\label{sect:eval}

\subsection{Datasets}

\textbf{Open source projects contributed by Samsung}: There are many kinds of open source projects contributed by Samsung Electronics such as Tizen, an open source operating system. Tizen runs on a wide range of Samsung devices including smartphones, tablets, in-vehicle infotainment devices, smart TVs, smart cameras, wearable computing (e.g. smartwatches such as Gear), smart home appliances (e.g. such as refrigerators, washing machines, air conditioners, and ovens/microwaves). We collected potential defects from those open source projects. To identify defective files, we employed a static analysis tool\footnote{Name is not revealed due to non-disclosure agreement.} used by Samsung that has specific support for target projects. This tool scans the source code of those projects and generates a report describing all the potential defects (i.e. warnings) that it can discover. There are different types and severity levels of warnings. In this study, we focused on critical resource leakage warnings (e.g. a handle was created but lost without releasing it.). We use these information to label files as defective or clean: a file is considered defective if the tool reported at least one resource leakage warning associated with that file. We built up a dataset of 8,118 files written in C, 2,887 of which (35.6\%) are labelled as defective and 5,231 (64.4\%) labelled as clean. \\

\hspace{-0.4cm}\textbf{PROMISE dataset:} We also used a dataset for defect prediction which is publicly available from the PROMISE data repository. To facilitate comparison, we selected the same 10 Java projects and release versions from this dataset as in \cite{Wang:2016:ALS}.  These projects cover a diversity of application domains such as XML parser, text editor, enterprise integration framework, and text search engine library (see Table \ref{table:dataset-stats}). The provided dataset only contained the project names, their release versions, and the file names and their defective labels. It did not have the source code for the files, which is needed for our study. Using the provided file names and version numbers, we then retrieved the relevant source files from the code repository of each application.

\begin{table}[h]
	\centering
	\caption{Dataset statistics}
	\label{table:dataset-stats}
	\resizebox{3.5in}{!}{%
		\begin{tabular}{@{}lcccccc@{}}
			\toprule
			App         & \#Versions     & \#Files    & Mean files       & Mean LOC   & Mean defective   & \% Defective    \\
			\midrule
			lucene & 3 & 750 & 250 & 47091 & 145 & 57.18 \\
			synapse & 3 & 635 & 211 & 30442 & 54 & 23.60 \\
			xerces & 2 & 891 & 445 & 132934 & 70 & 15.72 \\
			camel & 3 & 2379 & 793 & 81183 & 183 & 24.54 \\
			xalan & 2 & 1438 & 719 & 256625 & 248 & 33.53 \\
			ivy & 2 & 593 & 296 & 44288 & 28 & 9.00 \\
			ant & 3 & 1383 & 461 & 123452 & 96 & 19.88 \\
			jedit & 3 & 853 & 284 & 94696 & 81 & 28.85 \\
			poi & 3 & 1053 & 351 & 87611 & 223 & 63.14 \\
			log4j & 2 & 223 & 111 & 16979 & 35 & 32.07 \\
			\bottomrule
		\end{tabular}
	}
\end{table}


When processing the CSV spreadsheets provided with the PROMISE dataset, we have found that there were entries for inner classes. Since inner classes are included in an AST of their parent, we removed those entries from our dataset. We also removed entries for source files written in Scalar and entries that we could not retrieve the corresponding source files. In total, 264 entries were removed from the CSV spreadsheet.  Table \ref{table:dataset-stats} provides some descriptive statistics in our dataset.

\subsection{Performance measures}

Reporting the average of precision/recall across the two classes (defective and clean) is likely to overestimate the true performance, since our dataset is imbalance (i.e. the number of defective files are small). More importantly, predicting defective files is more of interest than predicting clean files. Hence, our evaluation is focus on the defective class.

A confusion matrix is  used to store the correct and incorrect decisions made by a prediction model. For example, if a file is classified as defetive when it is truly defective, the classification is a true positive (tp). If the file is classified as defective when it is actually clean, then the classification is a false positive (fp). If the file is classified as clean when it is in fact defective, then the classification is a false negative (fn). Finally, if the issue is classified as clean and it is in fact clean, then the classification is true negative (tn). The values stored in the confusion matrix are used to compute the widely-used Precision, Recall, and F-measure.

\begin{itemize}
	\item{Precision:} The ratio of correctly predicted defective files over all the files predicted as being defective. It is calculated as: \begin{displaymath}pr=\frac{tp}{tp+fp}\end{displaymath}

	\item{Recall:} The ratio of correctly predicted defective files over all of the true defective files. It is calculated as: \begin{displaymath}re=\frac{tp}{tp+fn}\end{displaymath}

	\item{F-measure:} Measures the weighted harmonic mean of the precision and recall. It is calculated as: \begin{displaymath}F-measure=\frac{2*pr*re}{pr+re}\end{displaymath}

	\item{Area Under the ROC Curve (AUC)} is used to evaluate the degree of discrimination achieved by the model. The value of AUC is ranged from 0 to 1 and random prediction has AUC of 0.5. The advantage of AUC is that it is insensitive to decision threshold like precision and recall. The higher AUC indicates a better prediction.
\end{itemize}

\subsection{Results}

\subsubsection{Within-project prediction} This experiment\footnote{All experiments were run on Intel(R) Xeon(R) CPU E5-2670 0 @ 2.6GHz. There machine has two CPUs, each has 8 physical cores or 16 threads, with a RAM of 128GB.} used data from the same project for both training and testing. For the Samsung dataset, we could not trace back which project a source file belonged to, and thus we treated all the source files in the dataset as belonging to a single project. We employed cross-fold validation and divided the files in this dataset into ten folds, each of which have the approximately same ratio between defective files and clean files. Each fold is used as the test set and the remaining folds are used for training. As a result, we built ten different prediction models and the performance indicators are averaged out of the ten folds. We also tested with two different classifiers: Random Forests and Logistic Regression.

\begin{figure}[h]
\centering \includegraphics[width=\linewidth]{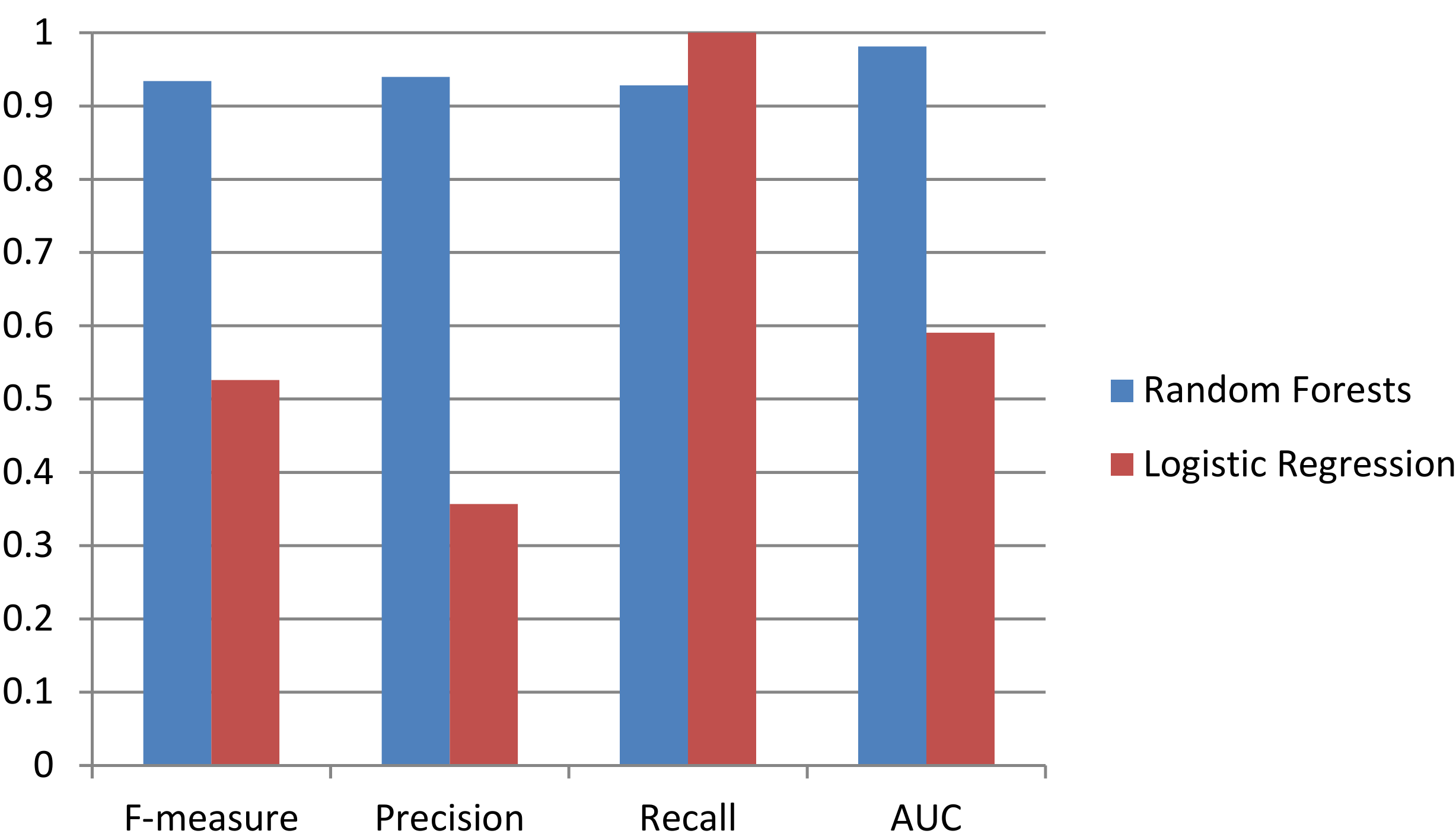}
\caption{Predictive performance of our approach for the Samsung dataset}
\label{fig:results-samsung-lstm}
\end{figure}

\begin{figure*}[ht]
\centering \includegraphics[width=\linewidth]{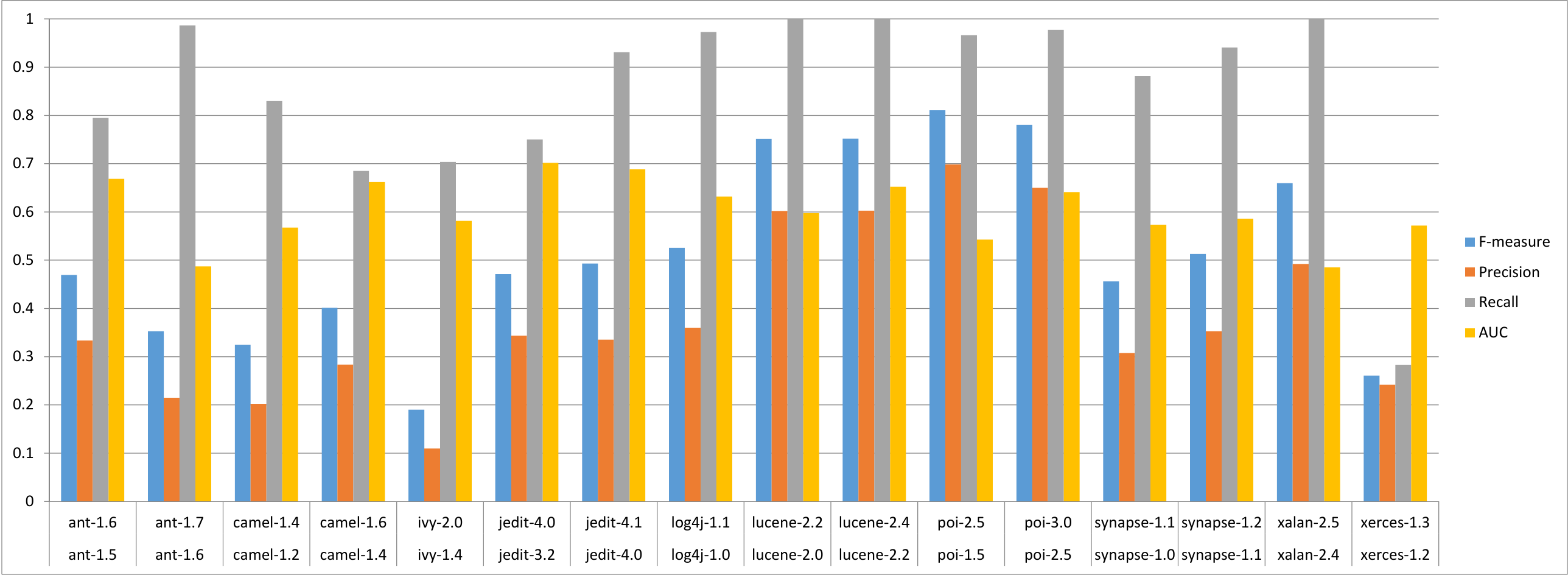}
\caption{Predictive performance of our approach for the Samsung dataset (within-project prediction). The X-axis has pairs of training (lower version) and testing data (the newer version) in each project. For example, in the first pair our model was trained using version 1.5 of Apache Ant project and tested using its version 1.6.}
\label{fig:results-samsung-lstm-within-project}
\end{figure*}

Figure \ref{fig:results-samsung-lstm} shows the predictive performance of our approach for the Samsung dataset. The predictive model which uses Random Forests (RF) as the classifier produced an impressive result with all four performance indicators (F-measure, Precision, Recall and AUC) being well above 0.9. Using Logistic Regression (LR) achieved very high recall, but at the same time it appeared to produce many false positives, and thus its precision is much lower than the precision produced by RF. Both classifiers achieved an AUC well above the 0.5 threshold (0.98 for RF and 0.60 for RF), suggesting that our approach is significantly better than random prediction.

For the PROMISE dataset, since it contains different versions of the same applications, we followed the setting in Wang \emph{et. al.} \cite{Wang:2016:ALS} and used two consecutive versions of each project for training and testing. Specifically, the source code of an older version is used to training the model and the later version is used for testing the model. In total, we conducted 16 sets of experiments exactly as in Wang \emph{et. al.}. We also tested with Random Forests and Logistic Regression as the classifier, and observed a different result (compared to the result for the Samsung dataset): using LR produced better predictive performance than using RF. This can be explained by the fact that the PROMISE dataset has small number of data points, which fits better with LR.

Due to space limitation, we reported here only the results from using LR as the classifier (see Figure \ref{fig:results-samsung-lstm-within-project}). Our prediction model produced an average AUC of 0.6, well above the random prediction threshold. More importantly, it achieved a very good recall of 0.86 (averaging across 16 cases), which is 23\% improvement over Wang \emph{et. al.}'s approach. However, our approach has lower precision, leading to a deduction in F-measure (17\%) compared against Wang \emph{et. al.}'s approach. We note that high recall is generally preferable in predicting defects since the cost of missing defects is much higher than having false positives.
 
\subsubsection{Cross-project prediction}

\begin{figure*}[ht]
\centering \includegraphics[width=\linewidth]{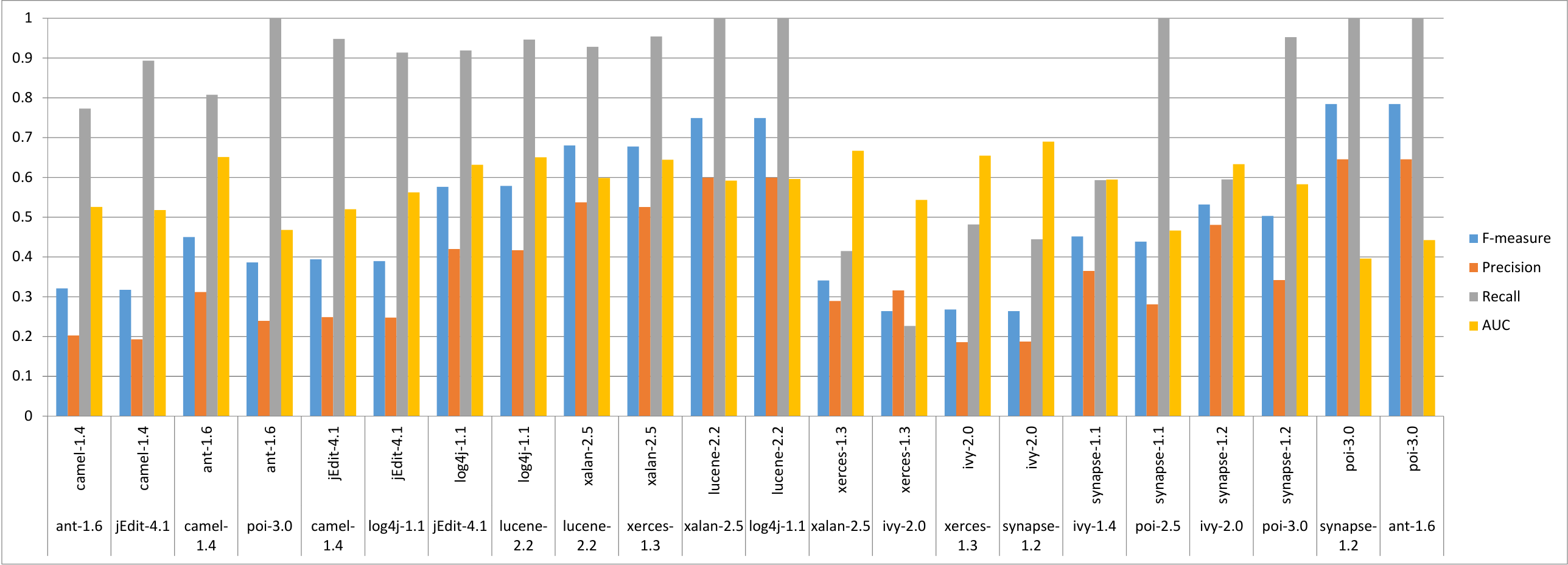}
\caption{Predictive performance of our approach for the Samsung dataset (cross-project prediction). The X-axis has pairs of training (source project) and testing data (target project). For example, in the first pair our model was trained using version 1.6 of Apache Ant project and tested using version 1.4 of the Camel project.}
\label{fig:results-samsung-lstm-cross-project}
\end{figure*}

Predicting defects in new projects is often difficult due to lack of training data. One common technique to address this problem is training a model using data from a (source) project, and applying it to the new (target) project.  We conducted this experiment by selecting one version from a project  in our PROMISE dataset as the source project (e.g. ant 1.6) and one version from another project as the target project (e.g. camel 1.4). Figure \ref{fig:results-samsung-lstm-cross-project} summarizes the results in cross-project prediction for the twenty-two pairs of source and target Java projects.

Our approach again achieved very high recall, with an average of 0.8 across 22 cases in cross-project prediction. There are 15 cases where the recall was above 0.8. The average F-measure is however 0.5, due to the low precision as seen in within-project prediction. However, the average AUC is still well above the 0.5 threshold, demonstrating the overall effectiveness of our approach in predicting defects.

\subsection{Threats to validity}\label{sect:threats}

There are a number of threats to the validity of our study, which we discuss below. We mitigated the construct validity concerns by evaluating our approach not just only on our internal dataset but also on a publicly available dataset (the PROMISE dataset). Both datasets contains real projects. The PROMISE dataset did not unfortunately contain the source files. However, we have carefully used the information (e.g. application details, version numbers and date) provided with the dataset to retrieve the relevant source files from the code repository of those applications. We tried to minimize threats to conclusion validity by using standard performance measures for defect prediction. We however acknowledge that a number of statistical tests \cite{STVR:STVR1486} can be applied to verify the statistical significance of our conclusions, which we plan to do in our future work.

With regard to internal validity, the Samsung dataset we used contains defective labels which were derived from warnings provided by a static analysis tool used internally at Samsung. We acknowledge that those warnings may contain false positives, and thus future work would involve investigating those warnings and confirming their validity. In addition, we did not have the source code to replicate Wang \emph{et. al.}'s experiments \cite{Wang:2016:ALS}, and thus had to rely on the results they reported to make a comparison with our approach. In terms of external validity. We have considered a large number of applications which differ significantly in programming language, size, complexity, domain, popularity and revision history. We however acknowledge that our data set may not be representative of all kinds of software applications. Further investigation to confirm our findings for other types of applications such as web applications and applications written in other programming languages such as PHP and C++.

    \section{Related Work}
\label{sect:related-work}

\subsection{Defect Prediction}
Research in defect prediction has faced multiple challenges in the past (e.g issues regarding the lack of availability, variety and granularity of data). Recent accomplishments from different researchers has made a huge impact in providing solutions to the different issues (e.g introduction of open source software).

Significant amount of research has been done in designing features to be used in defect prediction, which can be divided into static code features and process features. Static code features can be further broken down into code size and code complexity (e.g Halstead features, McAbe features, CK features, MOOD features). Process features measures the change activity in the development of a release in order to build more accurate defect prediction models. Motivation for the usage of process metrics arises from how different processes used in the software development may lead to defects. The usage of process metrics is independent of programming language, making it possible to be used in a wide range of projects. Models based on different machine learning techniques (e.g random forest), utilizing the  features as described before are evaluated within the same project or across different projects.

Within-project prediction uses data from the same project to build a model. Using this approach requires large amount of data in order for it to be effective. Zimmermann et al. \cite{4814164} proposed the usage of network measures in building a defect prediction model which has been evaluated to perform better than using complexity metrics. Specifically, network analysis was performed on the dependency graphs of Windows Server 2003. A different approach which has been widely used recently is to build a cross-project defect prediction model. Li et al \cite{8009936} proposes an approach for defect prediction using deep learning (i.e Convolutional neural network). The proposed framework called Defect Prediction via Convolutional Neural Network (DP-CNN), when evaluated performs better than existing approaches (e.g traditional, DBN) in defect prediction.

Cross-project prediction uses historical data from other projects to train the model. Zimmermann et al. \cite{Zimmermann:2009:CDP:1595696.1595713} evaluated 622 cross-project defect prediction models using 12 different applications. Building an accurate cross-project prediction model is difficult and overcoming this challenge is of great significance for instances where there is insufficient data to build a model. Zhang et al. \cite{Zhang:2014:TBU:2597073.2597078} builds a universal defect prediction model from different projects after the predictors are preprocessed using a context-aware rank transformation. The performance of the universal model is similar as compared to within-project predictions and also when tested against five other projects.

Defect prediction is a very active area in software analytics. Since defect prediction is a broad area, we highlight some of the major work here, and refer the readers to other comprehensive reviews (e.g. \cite{D'Ambros:2012:EDP,Catal:2009:SRS}) for more details. Code metrics were commonly used as features for building defect prediction models (e.g. \cite{Hall:2012}). Various other metrics have also been employed such as change-related metrics \cite{Moser:2008,Nagappan:2005:URC}, developer-related metrics \cite{Pinzger:2008:DNP}, organization metrics \cite{Nagappan:2008:IOS}, and change process metrics \cite{Hassan:2009:PFU}.

Recently, a number of approaches (e.g. \cite{Yang:2015:DLJ,Wang:2016:ALS}) have leveraged a deep learning model called Deep Belief Network (DBN) \cite{DBN-Hinton06} to automatically learn features for defect prediction and have demonstrated an improvement in predictive performance. In fact, according to the evaluation reported by Wang \emph{et. al.} \cite{Wang:2016:ALS} their DBN approach outperformed both the software metrics and Bag-of-Word approaches. DBN however does not naturally capture the sequential order and long-term dependencies in source code. Most of the studies in defect prediction operate at the file level. Recent approaches address this issue at the method level (e.g. \cite{Giger:2012:MBP}) and the line level (e.g. \cite{Ray:2016:NBC}). Since our approach is able to learn features at the code token level, it may work at those finer level of granularity. However, this would require the development of new datasets which contain methods and codelines with defect labels, which we leave for future work.


\subsection{Deep learning in code modeling}

Deep learning has recently attracted increasing interests in software engineering. In our recent vision paper \cite{DeepSoftFSE2016}, we have proposed DeepSoft, a generic deep learning framework based on LSTM for modeling both software and its development and evolution process. We have demonstrated how LSTM is leveraged to learn long-term temporal dependencies that occur in software evolution and how such deep learned patterns can be used to address a range of challenging software engineering problems ranging from requirements to maintenance. Our current work realizes one of those visions.

The work in \cite{White:2015:TDL} demonstrated the effectiveness of using recurrent neural networks (RNN) to model source code. Their later work \cite{White:2016:DLC} extended these RNN models for detecting code clones. The work in \cite{Gu:2016:DAL} uses a special RNN  Encoder--Decoder, which consists of an encoder RNN to process the input sequence and a  decoder  RNN  with  attention  to  generate  the  output  sequence, to generate  API  usage  sequences  for a  given  API-related  natural  language  query. The work in \cite{DBLP:conf/aaai/GuptaPKS17} also uses RNN Encoder--Decoder but for fixing common errors in C programs. The work in \cite{Huo:2016:LUF} uses Convolutional Neural Networks (CNN) \cite{Cun:1990:HDR} for bug localization. Preliminary results from our earlier work \cite{DeepCode2016} also suggest that LSTM is a more effective language model for source code. Our work in this paper also develops a representation for source code but we use Tree-LSTM to better match with the Abstract Syntax Tree representation of code.

    \section{Conclusions and future work}\label{sect:conclusions}

We have presented a novel approach to predict defects in source code. Our prediction model takes as input an Abstract Syntax Tree (AST) representing a source file, a common representation for source code, and predict if the file is defective or clean. Our prediction system is built upon the powerful deep learning Long Short-Term Memory (LSTM) architecture to capture the long-term dependencies which often exist between code elements. Our novel use of the tree-structured LSTM network (Tree-LSTM) naturally matches the AST representation, and thus sufficiently captures the syntax and different levels of semantics in source code. All the features used in our prediction system are automatically learned through training the Tree-LSTM model, thus eliminating the need for manual feature engineering which occupies most of the effort in traditional approaches. We performed an evaluation on two different datasets provided by Samsung and the PROMISE repository. Promising results from our evaluation demonstrate that our approach can be applied into practice. 

Our future work involves applying this approach to other types of applications (e.g. Web applications) and programming languages (e.g. PHP or C++). We also aim to extend our approach to predict defects at the method and code change levels. In addition, we plan to explore how our approach can be extended to predicting specific types of defects such as security vulnerability and safety-critical hazards in code. Finally, our future development also involves building our prediction model into a tool which can be used to support software engineers and testers in real-life settings. 

	
\section*{Acknowledgement}
The authors gratefully acknowledge support from Samsung through its 2016 Global Research Outreach Program.

	\bibliographystyle{ACM-Reference-Format}


\end{document}